\documentclass[
aps,pra,
reprint,
a4paper,
superscriptaddress,
longbibliography,
]{revtex4-1}
	\usepackage[utf8]{inputenc}
\usepackage[T1]{fontenc}
\usepackage{textcomp}

\usepackage{amsfonts}
\usepackage{amsmath}
\usepackage{amsthm}
\usepackage{braket}
\usepackage{graphicx}
\usepackage{hyperref}

\hypersetup{citecolor=magenta}
\hypersetup{colorlinks=true}
\hypersetup{linkcolor=blue}
\hypersetup{urlcolor=blue}

\DeclareMathOperator{\tr}{tr}
\providecommand{\reals}{{\mathbb R}}
\newcommand{\exv}[1]{{\langle{#1}\rangle}}

\begin{document}
\title{Memory cost for simulating all quantum correlations from the 
Peres–Mermin scenario}

\author{Gabriel Fagundes}
\email{gabrielf@fisica.ufmg.br}
\affiliation{Departamento de Física, Universidade Federal de Minas Gerais UFMG, 
P.O.~Box 702, 30123–970, Belo Horizonte, MG, Brazil}
\affiliation{Department of Theoretical Physics, University of the Basque 
Country UPV/EHU, P.O.~Box~644, 48080~Bilbao, Spain}

\author{Matthias Kleinmann}
\affiliation{Department of Theoretical Physics, University of the Basque 
Country UPV/EHU, P.O.~Box~644, 48080~Bilbao, Spain}

\begin{abstract}
Sequences of compatible quantum measurements can be contextual and any 
 simulation with a classical model conforming with the quantum predictions 
 needs to use internal memory.
Kleinmann \emph{et al.} [New\ J.\ Phys.\ \textbf{13}, 113011 (2011)] showed 
 that simulating the sequences from the Peres–Mermin scenario requires at least 
 three different internal states in order to be not in contradiction with the 
 deterministic predictions of quantum theory.
We extend this analysis to the probabilistic quantum predictions and ask how 
 much memory is required to simulate the correlations generated for sequences 
 of compatible observables by any quantum state.
We find that even in this comprehensive approach only three internal states are 
 required for a perfect simulation of the quantum correlations in the 
 Peres–Mermin scenario.
\end{abstract}

\maketitle

\section{Introduction}
In the standard formulation of quantum theory (QT) the individual outcomes of 
 measurements are, in general, not predetermined by the state of the system.
Consequently, QT allows us to asses only the probability distribution over the 
 measurement outcomes.
Specker \cite{Specker:1960DIA} noted that this is a fundamental property of QT 
 and if quantum measurements had predetermined outcomes it would imply that 
 these values depend on the measurement context.
In this sense, QT is contextual and the mathematical formulation of this 
 observation is the Kochen–Specker theorem \cite{Kochen:1967JMM}.

Significant effort has been undertaken to understand the connection between 
 quantum contextuality and quantum information theory, for example, with 
 respect to the advantage of quantum computing over classical computing 
 \cite{Raussendorf:2013PRA, Howard:2014NAT}.
Similarly, a quantum system distributed over several parties can be used to 
 reduce the communication complexity over what is possible with classical 
 systems alone \cite{Toner:2003PRL, Buhrman:2010RMP} and the communication 
 advantage has been identified as a resource \cite{Gallego:2012PRL, 
 deVicente:2014JPA, Barrett:2005PRA}.
A related concept is the memory cost in sequential measurements 
 \cite{Kleinmann:2011NJP, Brierley:2015PRL}, i.e., the memory needed to 
 simulate the correlations occurring in sequences of quantum measurements by 
 means of a classical automaton with memory.
It has been found that the memory cost can exceed the amount of information 
 that can be stored in the quantum system yielding a quantum memory advantage 
 \cite{Galvao:2003PRL, Kleinmann:2011NJP, Brierley:2015PRL}.
We are here interested in the analysis of the memory cost with respect to 
 quantum contextuality, i.e., to determine the memory cost when the 
 measurements in a sequence only embraces mutually compatible measurements 
 \cite{Kleinmann:2011NJP}.
In this strict form the question of whether there exists a quantum memory 
 advantage due to contextuality is still open.

In this paper we investigate the situation for one of the most natural 
 candidates for a quantum memory advantage, the Peres–Mermin square.
We ask, what is the smallest memory for a classical model to reproduce all 
 contextuality predictions from the Peres–Mermin scenario, for any quantum 
 state.
Our focus here is to stay strictly in the regime of quantum contextuality, 
 i.e., sequences of compatible measurements, and to take into account also the 
 probabilistic predictions of quantum theory, while at the same time to admit 
 the most versatile classical automaton models.

\section{The Peres–Mermin square}
A simple proof of the Kochen–Specker theorem was found by Peres 
 \cite{Peres:1990PLA} and Mermin \cite{Mermin:1990PRL} and uses 9 quantum 
 observables arranged in the Peres–Mermin square,
\begin{equation}
\begin{bmatrix}A&B&C\\a&b&c\\\alpha&\beta&\gamma\end{bmatrix}=
\begin{bmatrix}
 \sigma_z\otimes \openone&\openone \otimes\sigma_z &\sigma_z\otimes\sigma_z\\
 \openone\otimes\sigma_x&\sigma_x\otimes\openone &\sigma_x\otimes\sigma_x\\
 \sigma_z\otimes\sigma_x& \sigma_x\otimes\sigma_z &\sigma_y\otimes\sigma_y
\end{bmatrix},
\end{equation}
 where $\sigma_x$, $\sigma_y$, and $\sigma_z$ are the Pauli operators.
The proof of the theorem consists of the observations (i) that the operators 
 within each row and each column form a context, i.e., they are mutually 
 compatible, and (ii) that the condition
\begin{equation}
 ABC=abc=\alpha\beta\gamma=Aa\alpha=Bb\beta=-Cc\gamma=\openone
\end{equation}
 holds.
Therefore, according to QT, the expected value of the product of the outcomes 
 of observables in one context is always $+1$, with the exception 
 $\exv{Cc\gamma}= -1$.
In order to obtain this behavior if the values of the observables are 
 predetermined, at least one observable needs to have a context-dependent 
 value, so that, for example, $\gamma$ has value $+1$ in the context 
 $\alpha\beta\gamma$ but value $-1$ in the context $Cc\gamma$.

In QT, the outcomes of all observables within a context can be obtained in a 
 joint measurement.
For the three dichotomic observables in each context of the Peres–Mermin 
 square, the joint measurement on two qubits has four distinct outcomes, taken 
 from the set of the 8 possible combinations of outcomes $\set{ (+1,+1,+1), 
 (+1,+1,-1), \dotsc, (-1,-1,-1)}$.
Alternatively, the outcomes can be obtained by measuring the observables in a 
 context sequentially.
This approach has been preferred in recent experiments on quantum contextuality 
 \cite{Kirchmair:2009NAT, Amselem:2009PRL, Lapkiewicz:2011NAT, Zhang:2013PRL, 
 DAmbrosio:2013PRX, Jerger:2016NCOM}.
When an observable $X$ from the Peres–Mermin square is measured, then the 
 quantum state $\rho$ changes according to
\begin{equation}
 \rho\mapsto \frac{\Pi_{x|X}\rho\Pi_{x|X}}{\tr(\rho \Pi_{x|X})},
\end{equation}
 with $\Pi_{x|X}=\tfrac12(\openone+xX)$ depending on the measurement outcome 
 $x=\pm1$ of $X$.
In a sense, sequential measurements with this Lüders transformation 
 \cite{Luders:1951APL} are a special way to implement a joint measurement.
Since the quantum state changes according to the choice of the observable and 
 the measurement outcome, one can argue that the quantum state serves as a 
 memory and the contextual behavior is achieved due to the very presence of 
 this memory.

However, in an extended variant of the Peres–Mermin square, it has been shown 
 \cite{Kleinmann:2011NJP} that even if one takes this perspective, a classical 
 model mimicking the quantum behavior would need more than four internal 
 states.
This extended scenario uses quantum predictions for all combinations of Pauli 
 matrices on two qubits, resulting in 15 dichotomic observables.
The classical model must then reproduce the predictions from any sequence of 
 compatible observables as well as respect conditions of compatibility and 
 repeatability.
The latter include conditions on sequences of incompatible measurements, and 
 thus are outside the contextuality paradigm.
Since the extended variant also operates on a quantum four-level system and 
 such systems can carry at most two bits of classical information 
 \cite{Holevo:1973PIT}, this has been identified as an instance of memory 
 advantage \cite{Galvao:2003PRL, Kleinmann:2011NJP, Brierley:2015PRL}.

The analysis in Ref.~\cite{Kleinmann:2011NJP} concerns classical models which 
 reproduce the deterministic predictions of QT within a sequence.
Such predictions are, for example, that the product of outcomes in the sequence 
 $A,B,C$ is always $+1$ or that the value of $A$ is repeated in the sequence 
 $A,B,A$.
For the case of the Peres–Mermin square and when any sequence of measurements 
 consists of observables from one context, there is a classical model 
 consistent with QT in this sense and which only uses three internal states.
This analysis does not cover the probabilistic predictions of QT, for example, 
 that $\exv{A}= 0$ for certain quantum states and it is not known how much 
 memory is needed to reproduce also the probabilistic predictions of QT in the 
 Peres–Mermin square.
Since the Peres–Mermin scenario is tightly linked to contextuality, we only 
 consider sequences of observables taken from one context.
This includes predictions like $\exv{BBA}= \exv{A}$, but excludes predictions 
 involving incompatible observables as in $\exv{ABc\gamma}=-1$.
In this paper our aim is hence to determine the smallest memory for a classical 
 model to reproduce the nondeterministic contextual quantum predictions from 
 the Peres–Mermin scenario, for any quantum state.

\section{Sequential correlations and stochastic automata}
The outcomes of a sequence of quantum measurements may be viewed as an 
 input–output process operating on a quantum system.
The input is the choice of the observable $X$ and the output is the outcome $x$ 
 of the measurement of the observable.
The overall probability for an output sequence $x_1,x_2,\dotsc$ for a given 
 input sequence $X_1,X_2,\dotsc$ is $P(x_1,x_2,\dotsc|X_1,X_2,\dotsc)$ and 
 within standard QT only such correlations can be predicted.

\begin{figure}
\includegraphics[width=.95\linewidth]{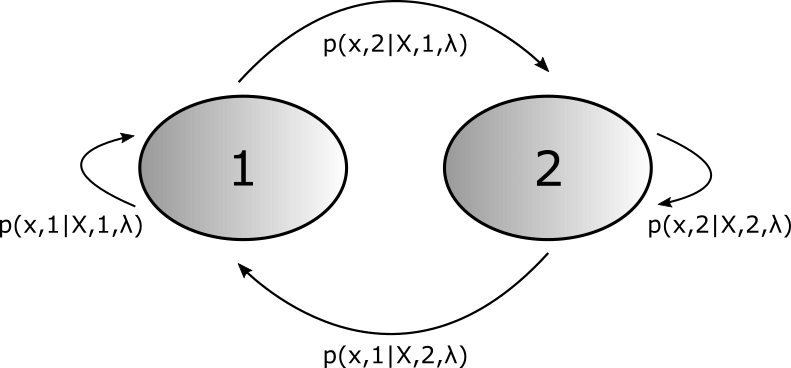}
\caption{\label{Fig1}%
Stochastic sequential machine with two internal states, $M= \set{1,2}$.
The transitions between the states $s$ and $s'$ are represented by arrows.
The probability $p(x,s'|X,s,\lambda)$ for the output $x$ and the transition 
 from state $s$ to state $s'$ can depend on the input $X$ and the external 
 parameter $\lambda$.
}
\end{figure}

The classical counterpart is modeled by an automaton which operates on 
 classical memory.
This memory is represented by a set $M$ of internal memory states.
In addition, the automaton has access to an external source of randomness, 
 modeled by an external parameter $\lambda$ which is fixed throughout a 
 measurement sequence but randomly distributed among different sequences 
 according to a distribution function $p(\lambda)$.
We use the model of a stochastic sequential automaton \cite{Paz:1971} where the 
 output $x$ and the state $s'\in M$ after the output only depend on the input 
 $X$, the value of $\lambda$, and the internal state $s\in M$ before the 
 output, cf.\ Fig~\ref{Fig1}.
The behavior of the automaton is hence summarized by the probability 
 distribution $p(x,s'|X,s,\lambda)$.
It represents the probability of the output $x$ and subsequent transition to 
 the internal state $s'$, given the input $X$, the current internal state $s$ 
 and the value of the parameter $\lambda$.
Similarly, the initial state of the automaton has a distribution depending on 
 $\lambda$, which we write as $p(s_0|\lambda)$.
With this model, the correlations achieved by the automaton are
\begin{multline}\label{eq:mealy}
P(x_1,x_2,\dotsc|X_1,X_2,\dotsc)=\\
 \sum_{\lambda,s_0,s_1,s_2,\dotsc}p(\lambda)p(s_0|\lambda)
 p(x_1,s_1|X_1,s_0,\lambda)\\
 \times p(x_2,s_2|X_2,s_1,\lambda) \dotsm.
\end{multline}
For a given automaton, i.e., $p(x,s'|X,s,\lambda)$ and $p(s_0|\lambda)$, the 
 choice of $p(\lambda)$ yields different correlations, so that the correlations 
 predicted by different quantum states can be reproduced using different 
 choices of the probability distribution $p(\lambda)$.

Clearly, it is possible to reproduce all noncontextual correlations with only 
 one internal state, $\lvert M\rvert= 1$, since in this case the right hand 
 side of Eq.~\eqref{eq:mealy} reduces to a hidden variable model 
 \cite{Bell:1964PHY}, $\sum_\lambda p(\lambda) p(x_1|X_1,\lambda) 
 p(x_2|X_2,\lambda) \dotsm$.
The external parameter $\lambda$ is not always included in such an analysis, 
 see, for example, the $\epsilon$-transducers studied in 
 Ref.~\cite{Barnett:2015JSP}.
However, then even noncontextual scenarios could require memory, since, for 
 example, measuring the sequence $\sigma_x,\sigma_x$ on an eigenstate of 
 $\sigma_z$ gives a random outcome for the first measurement, but the second 
 measurement has to repeat the value of the first measurement.
Consequently, if $\lambda$ does not occur, the simulation requires two internal 
 states, while when $\lambda$ can take two values, no memory is required.
The automaton is allowed to be intrinsically random, i.e., the distributions 
 $p(x,s'|X,s,\lambda)$ and $p(s_0|\lambda)$ may be nondeterministic.
As it is evident from our analysis below, this intrinsic randomness is not 
 required for simulating the quantum correlations from the Peres–Mermin 
 scenario.

\section{A memory-optimal automaton for the Peres–Mermin scenario}
As explained above, quantum contextuality is a feature of sets of compatible 
 observables and we therefore only consider sequences of measurements where all 
 observables are taken from one context.
That is, the observables are either taken from one row or one column of the 
 Peres–Mermin square.
Our first concern is the simulation of quantum measurements of a single 
 sequence of compatible observables.
According to QT, certain events can never occur, examples are the output 
 $+1,-1$ in the sequence $A,A$ or the output $+1,+1,-1$ in the sequence 
 $A,B,C$.
In Ref.~\cite{Kleinmann:2011NJP}, it has been shown that any automaton which 
 obeys all such quantum predictions must have memory with at least three 
 internal states.
An explicit example of such an automaton is given by \cite{Kleinmann:2011NJP}
\begin{equation}\label{eq:automaton1}\begin{split}
 o_1=\begin{bmatrix}+1&+1&+1\\+1&+1&+1\\+1&+1&+1\\\end{bmatrix},\quad &
 t_1=\begin{bmatrix}1&1&2\\1&1&3\\1&1&1\\\end{bmatrix},\\
 o_2=\begin{bmatrix}+1&+1&+1\\-1&+1&-1\\-1&-1&+1\\\end{bmatrix},\quad &
 t_2=\begin{bmatrix}2&1&2\\2&2&2\\2&3&2\\\end{bmatrix},\\
 o_3=\begin{bmatrix}+1&-1&-1\\+1&+1&+1\\-1&-1&+1\end{bmatrix},\quad &
 t_3=\begin{bmatrix}3&3&3\\1&3&3\\2&3&3\end{bmatrix}.
\end{split}\end{equation}
This notation is supposed to be read as follows.
Each matrix $o_i$, $i \in M$, represents the deterministic output for each of 
 the three internal states $M=\set{1,2,3}$.
Similarly, the transition matrices $t_i$ represent the internal state after the 
 output.
In terms of Eq.~\eqref{eq:mealy}, the distribution $p(x,s'|X,s)$ is $1$ if the 
 entry in the output matrix $o_s$ at the position of the observable $X$ is $x$ 
 and the entry in the transition matrix $t_s$ in the same position is $s'$; the 
 distribution is $0$ otherwise.
Here, $x\in \set{+1,-1}$, $s,s'\in M=\set{1,2,3}$, and $X\in 
 \set{A,B,C,a,b,c,\alpha,\beta,\gamma}$.
For example, if the automaton is in state $s=1$ and we provide the observable 
 $C$ as input, then the measurement outcome is $x=+1$ and the automaton changes 
 to the state $s'=2$.
It is straightforward to verify that this automaton obeys all deterministic 
 predictions of QT for any sequence of compatible observables 
 \cite{Kleinmann:2011NJP} and for any initial state $s_0$.

However, no quantum state gives deterministic predictions for all 9 observables 
 in the Peres–Mermin square, because these observables are not all mutually 
 compatible and no common eigenstate can exist.
In the following we extend the automaton from Eq.~\eqref{eq:automaton1} to use 
 the external parameter $\lambda$, so that a statistical mixture $p(\lambda)$ 
 can reproduce the quantum predictions.

\subsection{Other valid automata}
Starting from the automaton in Eq.~\eqref{eq:automaton1}, there are several 
 transformations which lead to other automata with the same properties.
First, it is possible to flip the signs for the output under the constraint 
 that for each row and each column there is either no flip of signs or there 
 are exactly two flips of signs.
This generates 15 additional automata.
Second, it is possible to make any permutation of the rows or a permutation of 
 the first and second column.
We restrict ourselves to the three permutations of rows which leave one row 
 unchanged and to the permutation of the first and second column.
This yields 4 additional automata and combined with the first set of 
 transformations we get in total $16\times5= 80$ automata.
In addition, we are free to choose the initial state $s_0$ and get this way 240 
 different behaviors.
We combine all these behaviors into a single automaton by allowing 240 
 different values for $\lambda$, i.e., the value of $\lambda$ determines the 
 behavior of the automaton.

\subsection{Example: Singlet state}\label{s:singlet}
As an example, we reproduce all quantum correlations for the singlet state (the 
 quantum state yielding $\exv{C}= \exv{c}=\exv{\gamma}= -1$) by choosing a 
 distribution $p(\lambda)$ for $\lambda=1,2,\dotsc,240$.
We choose $p(\lambda)= \tfrac14$ if $\lambda \in 
 \set{\lambda_1,\lambda_2,\lambda_3,\lambda_4}$ and $p(\lambda)=0$ else.
For $\lambda_k$, $k=1,2,3$, the transition matrices $t_1^{(k)}$, $t_2^{(k)}$, 
 $t_3^{(k)}$ are as in Eq.~\eqref{eq:automaton1} and the outcome matrices 
 $o_s^{(k)}$ are given by
\begin{subequations}
\begin{equation}\begin{split}
 o_1^{(1)}=\begin{bmatrix}-1&+1&-1\\-1&-1&+1\\+1&-1&-1\\\end{bmatrix},~
 o_2^{(1)}=\begin{bmatrix}-1&+1&-1\\+1&-1&-1\\-1&+1&-1\\\end{bmatrix},\\
 o_3^{(1)}=\begin{bmatrix}-1&-1&+1\\-1&-1&+1\\-1&+1&-1\\\end{bmatrix},\\
\end{split}\end{equation}
\begin{equation}\begin{split}
 o_1^{(2)}=\begin{bmatrix}-1&+1&-1\\+1&+1&+1\\-1&+1&-1\\\end{bmatrix},~
 o_2^{(2)}=\begin{bmatrix}-1&+1&-1\\-1&+1&-1\\+1&-1&-1\\\end{bmatrix},\\
 o_3^{(2)}=\begin{bmatrix}-1&-1&+1\\+1&+1&+1\\+1&-1&-1\\\end{bmatrix},\\
\end{split}\end{equation}
\begin{equation}\begin{split}
 o_1^{(3)}=\begin{bmatrix}+1&-1&-1\\-1&-1&+1\\-1&+1&-1\\\end{bmatrix},~
 o_2^{(3)}=\begin{bmatrix}+1&-1&-1\\+1&-1&-1\\+1&-1&-1\\\end{bmatrix},\\
 o_3^{(3)}=\begin{bmatrix}+1&+1&+1\\-1&-1&+1\\+1&-1&-1\\\end{bmatrix},\\
\end{split}\end{equation}
\begin{equation}\begin{split}
 o_1^{(4)}=\begin{bmatrix}+1&-1&-1\\+1&+1&+1\\+1&-1&-1\\\end{bmatrix},~
 o_2^{(4)}=\begin{bmatrix}+1&-1&-1\\-1&+1&-1\\-1&+1&-1\\\end{bmatrix},\\
 o_3^{(4)}=\begin{bmatrix}+1&+1&+1\\+1&+1&+1\\-1&+1&-1\\\end{bmatrix}.
\end{split}\end{equation}
\end{subequations}
The initial state for all four cases is $s_0=2$, i.e., we have 
 $p(s_0|\lambda)=1$ if $s_0=2$ and zero else.
In principle one can now verify that for sequences of compatible observables, 
 all quantum correlations from the singlet state are indeed reproduced.
However, there is an infinite number of input sequences which needs to be 
 considered and it is our next step to reduce the number of sequences to a 
 finite set.

\subsection{A finite set of sufficient input sequences}
We show in this section that a finite number of input sequences suffices to 
 determine all correlations for all sequences.
Since we only consider sequences of observables from one context, as soon as 
 two different observables occur in a sequence, it is already possible to 
 predict the reminder of the sequence from the outcome of these two 
 observables.
This is because the product of outcomes of the three observables of each 
 context is always $+1$ or $-1$, depending on the context, and due to the 
 requirement that repeated occurrences of an observable in a sequence produce 
 repeated values.

Hence, it remains to consider sequences where initially one observable is 
 measured repeatedly, for example, $X,X,Y$.
In quantum mechanics we have
\begin{multline}
 P(x,x,\dotsc,x, y|X,X,\dotsc,X, Y)=\\
 \tr(\Pi_{y|Y}\Pi_{x|X}\rho\Pi_{x|X}\Pi_{y|Y})= P(x,y|X,Y),
\end{multline}
 for any number of repetitions of the input $X$ and output $x$.
However, for the automaton model we could have different values for outcome $y$ 
 in the sequences $X,X,\dots,X, Y$, depending on the number of repetitions of 
 $x$, since the value of $Y$ does not need to be fixed until $Y$ is actually 
 measured.
Thus, we have to consider how our specific model behaves in this situation.
For any value of $\lambda$, the behavior of our automaton is analogous to the 
 automaton in Eq.~\eqref{eq:automaton1} and for this automaton one observes 
 that the internal state $s'$ after an $\ell$-fold measurement of $X$ does not 
 depend on $\ell$, if $\ell\ge 1$.
Hence, the outcome of $X,X,\dotsc,X,Y$ is $x,x,\dotsc,x,y$ if and only if $X,Y$ 
 has outcome $x,y$.

In summary, our automaton with any choice of $p(\lambda)$ reproduces the 
 quantum correlations for a state $\rho$ for all sequences of compatible 
 observables, if and only if it does so for all sequences of length two.
For practical reasons, instead of dealing with the correlations $P(x,y|X,Y)$ we 
 use the equivalent set of expectation values
\begin{subequations}\begin{align}
 \exv X&=\sum_{x,y} x P(x,y|X,Y),\\
 \exv{XYX}&= \sum_{x,y} yP(x,y|X,Y), \text{ and}\\
 \exv{XY}&=\sum_{x,y} xy P(x,y|X,Y),
\end{align}\end{subequations}
 where in the second equation we used that the value of $X$ in the first and in 
 the last position are the same.
Note, that while in QT, we always have $\exv{XYX} = \exv Y$, this does not hold 
 for all ensembles $p(\lambda)$ in our automaton, as, for example, in 
 Eq.~\eqref{eq:automaton1} with initial internal state $s=1$, we have $\exv{c} 
 = 1$, but $\exv{CcC}= -1$.
However, we observe that $\exv{XY}= \exv{XY}$ for all $p(\lambda)$ and all 
 compatible $X$ and $Y$, a relation that also holds in QT for any state.

Therefore, we have to take into account 9 values $\exv X$, 18 values 
 $\exv{XY}$, and 36 values $\exv{XYX}$.
We enumerate these values by $j= 1, \dotsc, 63$ and collect for each $j$ the 
 values for all 240 values of $\lambda$ in a vector $\vec v_j$.
Then the expectation values $\vec q=(q_1,\dotsc,q_{63})$ can be achieved if and 
 only if $q_j= \vec v_j\cdot \vec p$ for some probabilities $\vec p$ with 
 $p_\lambda\equiv p(\lambda)$.
The set of achievable expectation values $\vec q$ is hence given by the 
 polytope
\begin{equation}\label{v-rep}
 P = \set{ \vec q | q_j=\vec v_j\cdot \vec p
           \text{ for all } j \text{ and some } \vec p }.
\end{equation}
Similarly, for the quantum correlations we have $63$ hermitian operators $Z_j$, 
 such that the expectation values $\vec q$ can be attained according to QT if 
 and only if $q_j=\tr(\rho Z_j)$ for all $j$ and some quantum state $\rho$.
The set of achievable expectation values $\vec q$ according to QT is 
 consequently the convex set
\begin{equation}
 Q = \set{\vec q | q_j = \tr(\rho Z_j)
          \text{ for all } j \text{ and some } \rho}.
\end{equation}

This allows us to easily verify the correctness of the example in 
 Sec.~\ref{s:singlet}, by comparing $\vec v_j\cdot \vec p$ with $\tr(\rho Z_j)$ 
 for all $j$ and for any quantum state $\rho$, finding a corresponding 
 distribution $p(\lambda)$ reduces to find probabilities $\vec p$ with $\vec 
 v_j\cdot\vec p= \tr(\rho Z_j)$ for all $j$.
This can be solved by means of linear programming and was in fact our method to 
 find $p(\lambda)$ for the singlet state in Sec.~\ref{s:singlet}.

\subsection{Simulation of the correlations of any quantum state}
\label{sec:mr}
We are now equipped with the necessary tools to prove that the correlations 
 of any quantum state can be simulated with a construction analogous to the 
 one in Sec.~\ref{s:singlet}.
According to our previous analysis, the question whether the quantum 
 predictions can be simulated by an appropriate distribution $p(\lambda)$ 
 reduces to the question whether the convex set $Q$ is contained in the 
 polytope $P$.
In order to make this question tractable, we use an equivalent representation 
 of the polytope, where it is written as a finite intersection of half-spaces 
 \cite{Grunbaum:2003} parametrized by vectors $\vec h_\ell$ and numbers 
 $\alpha_\ell$, so that
\begin{equation}\label{h-rep}
 P = \set{ \vec q | \vec h_\ell\cdot \vec q \le \alpha_\ell
           \text{ for all $\ell$} }.
\end{equation}
Using this half-space representation, $P$ contains $Q$ if and only if $\vec 
 h_\ell \cdot \tr(\rho \vec Z) \le \alpha_\ell$ for all $\ell$ and all $\rho$.
By writing
\begin{equation}\label{eqW}
 W_\ell= \alpha_\ell\openone- \vec h_\ell\cdot \vec Z,
\end{equation}
 this further simplifies to $\tr(\rho W_\ell)\ge 0$ for all $\ell$ and all 
 $\rho$.
That is, $Q\subset P$ holds if and only if all $W_\ell$ are positive 
 semidefinite.
Conversely, if we find a state with $\tr(\rho W_\ell)<0$ for some $\ell$, and 
 hence $W_\ell$ is not positive semidefinite, then our automaton cannot 
 simulate all quantum predictions for this state.

In principle, this can be tested directly.
However, since the polytope $P$ is given in the form of Eq.~\eqref{v-rep}, we 
 need to compute the half-space representation in Eq.~\eqref{h-rep}.
This can be achieved by using the Fourier–Motzkin elimination, but is known to 
 be a computationally hard task and for our problem we were not able to find a 
 direct solution.
The central observation to solve the problem nonetheless is that $Q$ spans a 
 rather low-dimensional affine space.
In particular, $Q$ is contained in the affine space $\vec a+U\equiv \set{\vec a 
 +\vec u | \vec u\in U}$, where $a_j= \tr(\rho Z_j)$ for some fixed $\rho_0$ 
 (for example, $\rho_0= \tfrac14\openone$) and $U$ is the linear space $U=\set{ 
 \vec u | u_j = \tr(G Z_j) \text{ for some $G$} }$ with $G$ any hermitian 
 operator obeying $\tr(\rho_0 G)=0$.
This holds true since we can always write $\rho=\rho_0+G$ for some $G$.
The dimension of the linear space $U$ is only $\dim U=9$, as can be found by 
 using the linear independence relations of the operators $Z_j$.
Therefore, $Q\subset P$ is equivalent to $Q\subset P\cap (\vec a+U)$ and our 
 problem reduces to calculate a half-space representation for the polytope 
 $P\cap (\vec a+U)$.
This problem is easily tractable, as we discuss in Appendix~\ref{app:cone}.
We obtain 24 nonzero operators $W_\ell$, each of which is positive 
 semidefinite.
This proves $Q\subset P$ and thus our automaton can simulate the quantum 
 correlations for any quantum state.
We mention that the nonzero operators $W_\ell$ are, up to an arbitrary positive 
 factor, exactly those 24 projectors of unit rank which commute with all 
 observables from one out of the six contexts in the Peres–Mermin square.

\section{Conclusions}
Quantum contextuality is considered as one of the key differences between the 
 microscopic world and the world governed by classical mechanics.
Recent experimental demonstrations of this phenomenon proceed by measuring 
 sequences of observables and yield a contradiction to the assumption of 
 noncontextuality, i.e., the assumption that the value of an observable does 
 not depend of which other compatible observables are measured alongside.
We revisited this conclusion for the case of the Peres–Mermin scenario in the 
 light of classical models which utilize internal memory in order to reproduce 
 the quantum behavior.
We showed that for this scenario an automaton using only three internal states 
 can reproduce the quantum correlations from any quantum state for any sequence 
 of compatible observables.
This model is also optimal, since a lower bound of three internal states was 
 already established \cite{Kleinmann:2011NJP}.
The memory cost of the Peres–Mermin scenario is therefore actually lower than 
 the canonical quantum implementation, which requires two qubits.
Since for quantum correlations involving sequences of incompatible observables, 
 the memory cost can also be larger than the memory of the quantum system, this 
 leaves open the question, whether there can be a quantum memory advantage when 
 restricted to sequences of compatible observables and if so, for which 
 contextuality scenario this occurs.

\begin{acknowledgments}
We thank
Costantino Budroni,
Adán Cabello,
Marcelo Terra Cunha,
Jan-Åke Larsson,
Marco Túlio Quintino, and
Géza Tóth,
for discussions.
This work was supported by CNPq, Conselho Nacional de Desenvolvimento 
Científico e Tecnológico, Brazil,
the FQXi Large Grant ``The Observer Observed: A Bayesian Route to the 
Reconstruction of Quantum Theory'',
the EU (ERC Starting Grant GEDENTQOPT), and by
the DFG (Forschungsstipendium KL 2726/2–1).
\end{acknowledgments}

\appendix
\section{Low-dimensional section of a polyhedral cone}\label{app:cone}
A central step in Sec.~\ref{sec:mr} is to compute the half-space representation 
 of the polytope $P\cap (\vec a+ U)$, where $P$ is a polytope, $\vec a\in P$ is 
 a vector and $U$ is a linear subspace of low dimension.

We first consider the equivalent problem for a polyhedral cone $\mathcal P= 
 \set{ A\vec r | \vec r \succeq 0 }$, where $\vec r\succeq0$ abbreviates 
 $r_k\ge 0$ for all $k$ and $A$ is some matrix with real entries.
For a matrix $K$, let $F$ be a matrix the range of which is the kernel of $KA$.
We have
\begin{equation}\label{trick}\begin{split}
 \mathcal P\cap \ker(K)
 &=\set{ A\vec r | \vec r\succeq 0\text{, } K(A\vec r) = 0 }\\
 &=\set{ A\vec r | \vec r\succeq 0\text{, } \vec r= F\vec s
         \text{ for some } \vec s}\\
 &=\set{ AF \vec s | F \vec s \succeq 0 }\\
 &=AF\set{ \vec s | F \vec s \succeq 0 },
\end{split}\end{equation}
 where we used that $KA\vec r=0$ implies $\vec r=F\vec s$ for some $\vec s$ 
 and, conversely, $(KA)F\vec s=0$ for any $\vec s$.
It follows that if we can obtain a matrix $F'$, such that $\set{\vec s | F\vec 
 s \succeq 0 }= \set{F'\vec s | \vec s\succeq 0}$, then $\mathcal P\cap \ker(K) 
 = \set{ AFF' \vec s | \vec s \succeq 0}$.

For our case, we extend the polytope $P$ from Eq.~\eqref{v-rep} to a polyhedral 
 cone $\mathcal P$ by adding $\vec e= (1,1,\dotsc,1)$ to the vectors $\vec v_j$ 
 and by dropping the constraint $\sum_i p_i= 1$, i.e., $\mathcal P= \set{ A\vec 
 r | \vec r\succeq 0 }$ and $A$ is the matrix with rows 
 $[e,v_1,\dotsc,v_{63}]$.
Then $(1,\vec q)\in \mathcal P$ if and only if $\vec q\in P$.
Similarly, we define the linear subspace $\mathcal U=\set{ (\lambda,\lambda 
 \vec a+\vec u) | \lambda\in \reals \text{ and } \vec u\in U}$, so that 
 $(1,\vec x)\in \mathcal U$ is equivalent to $\vec x\in \vec a+U$.

In order to apply Eq.~\eqref{trick}, we choose some matrix $K$ such that 
 $\ker(K)= \mathcal U$ and some matrix $F$ with range $\ker(KA)$.
Despite $F^T$ being a larger matrix than $A$, we find that $F'$ is rather easy 
 to compute.
The matrix $B=AFF'$ is then only of rank $\dim(\mathcal U)=10$ and a matrix 
 $B'$ with $\set{B \vec s | \vec s\succeq 0} = \set{ \vec y | B'\vec y\succeq 0 
 }$ can be computed at an instance.
We use the software \texttt{cddlib} \cite{cddlib} to generate the matrices $F'$ 
 and $B'$ and \texttt{iml} \cite{iml} to compute $K$ and $F$.
Both packages work with unlimited exact integer arithmetic and hence our 
 computation of $B'$ is exact.
We verify independently our results by using \texttt{porta} \cite{porta} to 
 compute $K$, $F$, $F'$ and $B'$.

Finally, we have that $\vec q\in P$ and $\vec q\in \vec a+U$ if and only if 
 $(1,\vec q)\in \mathcal P\cap \mathcal U$, i.e., if and only if $B'_{\ell,1}+ 
 \sum_j B'_{\ell,j+1} q_j\ge 0$ for all $\ell$.
Therefore, the operators $W_\ell$ defined in Eq.~\eqref{eqW} are given by
\begin{equation}
 W_\ell= B'_{\ell,1}\openone- \sum_j B'_{\ell,j+1} Z_j.
\end{equation}
As we showed in the main text, $Q\subset P\cap(\vec a+U)$ is equivalent to all 
 $W_\ell$ being positive semidefinite.
In our analysis, all operators $W_\ell$ satisfy this condition.

\bibliography{local}

\end{document}